\documentclass{article}

\usepackage[position, final]{neurips_2025}
\makeatletter
\newcommand{\@notice}{}
\makeatother
\setcitestyle{numbers}
\usepackage{graphicx}

\usepackage[utf8]{inputenc} 
\usepackage[T1]{fontenc}    
\usepackage{hyperref}       
\usepackage{url}            
\usepackage{booktabs}       
\usepackage{amsfonts}       
\usepackage{nicefrac}       
\usepackage{microtype}      
\usepackage{xcolor}         
\usepackage{subcaption}

\title{Dyadic Neural Dynamics: Extending Representation Learning to Social Neuroscience}

\author{%
  Mariya Glushanina\\
  Cognitive Science\\
  \'{E}cole normale sup\'{e}rieure\\
  Paris, France  \\
  \texttt{mariya.glushanina@gmail.com} \\
   \And
  Jeffrey Huang \\
  Department of Computer Science \\
  Columbia University, USA\\
  New York, NY 10027 \\
   \texttt{jh4797@columbia.edu} \\
  \AND
  Michelle McCleod \\
 University of Alabama, USA \\
  Tuscaloosa, AL 35401 \\
   \texttt{mlmccleod@crimson.ua.edu} \\
   \And
  Brendan Ames \\
  University of Southampton, UK \\
  Southampton, SO17 1BJ \\
   \texttt{B.Ames@soton.ac.uk} \\
   \And
   Evie A. Malaia \\
  University of Alabama, USA \\
  Tuscaloosa, AL 35401 \\
  \texttt{eamalaia@ua.edu} \\
}

\begin{document}

\maketitle

\begin{abstract}
  Social communication fundamentally involves at least two interacting brains, creating a unique modeling problem. We present the first application of Contrastive Embedding for Behavioral and Neural Analysis (CEBRA) to dyadic EEG hyperscanning data, extending modeling paradigms to interpersonal neural dynamics. Using structured social interactions between participants, we demonstrate that CEBRA can learn meaningful representations of joint neural activity that captures individual roles (speaker-listener) and other behavioral metrics (autism quotient/AQ-10, etc.). Our approach to characterizing interactions, as opposed to individual neural responses to stimuli, addresses the key principles of foundational model development — scalability and  cross-subject generalization —while opening new directions for representation learning in social neuroscience and clinical applications.
\end{abstract}

\section{Introduction}
Foundation models have advanced biosignal analysis by developing generalizable representations across subjects, tasks, and modalities. However, current approaches focus exclusively on individual participants, overlooking the inherently social nature of human cognition and behavior. Social interactions involve complex coordination between multiple neural systems, creating rich temporal dynamics that cannot be captured by analyzing individuals in isolation \cite{hamilton2021hyperscanning}. In this field, hyperscanning (simultaneous neural recording from multiple participants) is the appropriate method to capture interpersonal brain dynamics during real-world social interactions. Existing analyses of hyperscanned data rely primarily on predefined connectivity measures or frequency-domain approaches that may miss crucial temporal patterns and cross-brain relationships. 
CEBRA is a promising solution for learning behaviorally-relevant latent representations through contrastive learning on continuous neural data \cite{schneider2023learnable}. The ability of CEBRA to capture complex temporal dynamics while maintaining interpretability makes it well-suited for the challenges of dyadic neural analysis. By extending CEBRA to hyperscanning data, we can develop foundation models that capture the joint neural substrates of social interaction, especially in cases with clinical and pediatric populations.
This work implements CEBRA approach to dyadic interactions between participants with varying autism spectrum traits. This application is particularly valuable because recent research challenges traditional deficit models of autism, suggesting that communication difficulties arise from neurotype mismatches rather than inherent deficits \cite{crompton2025information}. The "double empathy problem" framework posits that autistic individuals communicate effectively with each other, while mixed-neurotype pairs show selective breakdown—a phenomenon requiring dyadic-level analysis.
This work establishes a foundation for scaling representation learning approaches to multi-participant neural recordings, with implications for understanding social cognition, developing brain-computer interfaces for social contexts, and advancing computational approaches to psychiatric disorders.

\section{Methods}

We recruited dyads (N = 30 pairs; 8 analyzed in this work) with varying autism quotient (AQ) scores to create a spectrum of neurotype combinations - approved by the University of Alabama IRB. This design enables analysis of how neurotype matching influences joint neural representations. Participants completed validated assessments including the Autism Quotient (AQ-10, \cite{lundin2019aq10}) to quantify individual differences in social communication traits and anxiety.
Participants engaged in structured face-to-face interactions during simultaneous EEG recording. The protocol consisted of speaker recounting two personally relevant stories (~6 minutes total), while listerner offered empathy and support; participant roles were then reversed. Post-interaction ratings assessed subjective engagement and emotional resonance.
Continuous EEG was recorded using 64-channel EGI HydroCel Geodesic Sensor nets (1000 Hz sampling, vertex reference, impedances <50 $k\Omega$). Our preprocessing pipeline prioritized reproducibility and artifact minimization: 1) Band-pass filtering (0.1-45 Hz) with specified rolloff characteristics; 2) downsampling to 250 Hz for computational efficiency; 3) manual artifact rejection across both participants (same segments were removed); 4) automated bad channel detection and interpolation (MNE-Python); 5) Extended-Infomax ICA (15 components) for physiological artifact removal. 

\textbf{Data cleaning: } Our preprocessing pipeline prioritized reproducibility and artifact minimization: 1) Band-pass filtering (0.1-45 Hz) with specified rolloff characteristics; 2) downsampling to 250 Hz for computational efficiency; 3) manual artifact rejection across both participants (same segments were removed); 4) automated bad channel detection and interpolation (MNE-Python); 5) Extended-Infomax ICA (30 components) for physiological artifact removal. 

\textbf{CEBRA mapping: }We applied the Contrastive Embedding for Behavioral and neural Representation Analysis (CEBRA) framework to map high-dimensional EEG time series into a low-dimensional latent space. CEBRA learns an embedding by bringing temporally or behaviorally similar neural activity patterns closer together while pushing dissimilar patterns apart, using a contrastive loss with time-offset self-supervision and, in some runs, behavioral labels. We trained models on z-normalized, channel-wise min--max--scaled EEG from paired participants, using 3-dimensional output embeddings and cosine distance.

CEBRA setup: 
\begin{verbatim}
    MODEL_KWARGS = dict(
    model_architecture="offset10-model",
    batch_size=512,
    learning_rate=3e-4,
    temperature_mode = 'auto',
    temperature=1,
    min_temperature = 1e-1,
    max_iterations=5000,
    conditional="time_delta",
    output_dimension=3,
    distance="cosine",
    device="cuda:0",
    verbose=True,
    time_offsets=10,
)
\end{verbatim}

To assess mapping quality, we extracted four primary evaluation metrics from the trained models:
1)  Goodness-of-fit score --- the proportion of variance in behavioral labels explained by the learned embeddings (computed on evaluation data); 2) Decoding accuracy (KNN) --- classification accuracy of a 5-nearest-neighbor decoder trained on the embedding to predict the original labels; 3)  InfoLoss --- the final value of the contrastive InfoNCE loss, indicating how well the model separates dissimilar samples while pulling similar ones together; 4) Consistency scores between runs --- a measure of embedding stability, computed as the average similarity between embeddings obtained from independent training runs with different random seeds.

These metrics quantify how well the embedding preserves behaviorally relevant information and allow comparison across runs, label schemes, and control conditions (neural shuffle, behavioral shuffle, cross-shift, and cut-check).

\section{Results}

In order to receive meaningful embeddings, we had to ensure that the results were meaningful and were not subject to randomness, preprocessing artifacts or any other noise unrelated to the actual neural representations. 

We introduce three controls or sanity checks for our CEBRA-based procedure. In the first two, we replicate our analysis the after neural and label shuffling of the data, respectively. In each case, we observe non-meaningful embeddings and noisy loss; this is to be expected as the control process eliminates correlation between labels and the observed EEG signals.

\begin{figure}[h]
    \centering
    \begin{subfigure}{0.45\linewidth}
        \centering
        \includegraphics[width=\linewidth]{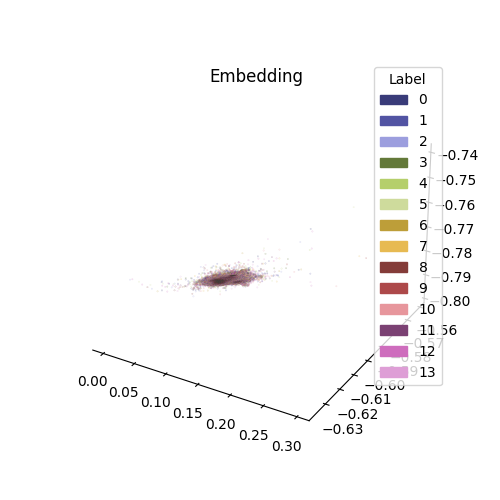}
        \caption{Behavioral control embedding from V1 labels run} 
    \end{subfigure}
    \hfill
    \begin{subfigure}{0.45\linewidth}
        \centering
        \includegraphics[width=\linewidth]{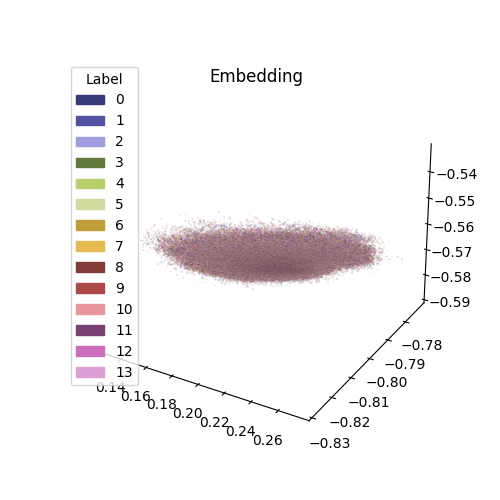}
        \caption{Neural control embedding from V1 labels run}
    \end{subfigure}
    \hfill
    \begin{subfigure}{0.45\linewidth}
        \centering
        \includegraphics[width=\linewidth]{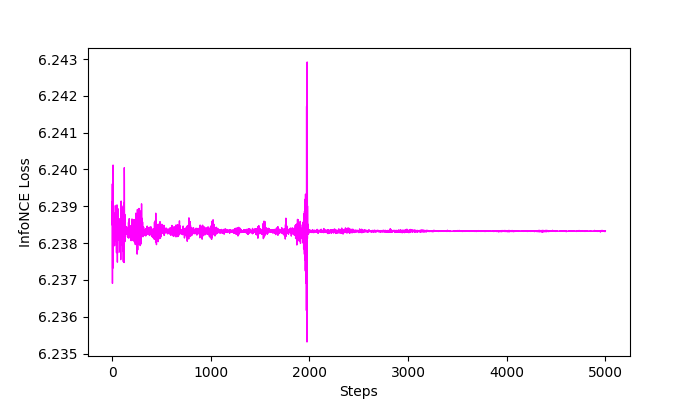}
        \caption{Behavioral control loss from V1 labels run}
    \end{subfigure}
        \hfill
    \begin{subfigure}{0.45\linewidth}
        \centering
        \includegraphics[width=\linewidth]{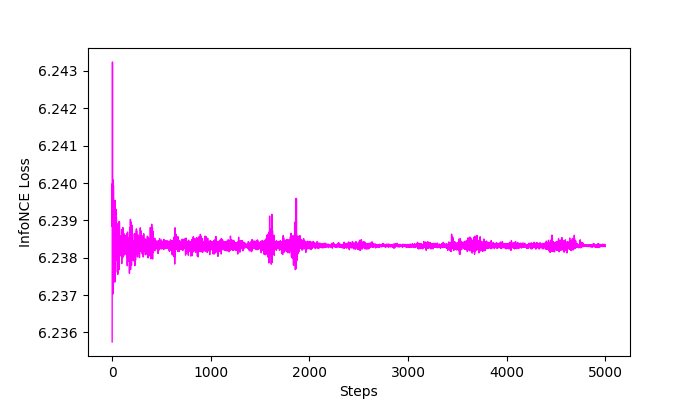}
        \caption{Neural control loss from V1 labels run}
    \end{subfigure}
    \caption{Behavioral and neural control embeddings and labels from V1 labels run}
    \label{fig:v1_controls7}
\end{figure}

Another sanity check related to our preprocessing procedure. As we performed the manual cutting of the bad data segments, we also aligned them, which resulted in specific artifacts after concatenation. To ensure, that CEBRA learnt meaninful embeddings instead of this artifacts of concatenation we introduced another check: we introduced random 50ms cuts to both files in the dyad. One dyad received 5 extra cuts, another dyad received N of cuts equivalent to the number of actual cuts, that overlapped with the ones present, to shift their position + 5 extra cuts in the random spots. This ensured randomness in concatenation between dyads. The results show no significant bias from the data with the aligned cuts, suggesting the algorithm learns meaningful representations in all four types of labels. \\  

\begin{figure}[h]
    \centering
    \begin{subfigure}{0.45\linewidth}
        \centering
        \includegraphics[width=\linewidth]{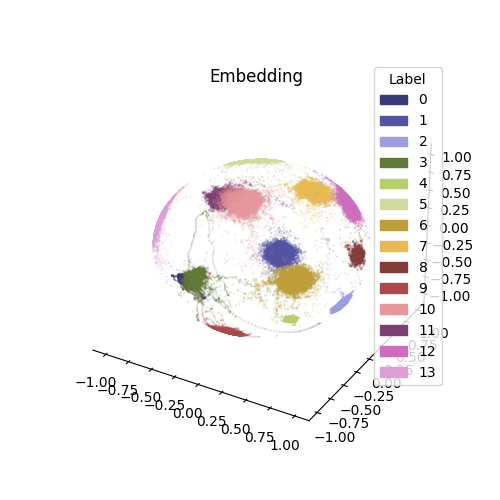}
        \caption{Pseudocut control embedding from V1 labels run} 
    \end{subfigure}
    \hfill
    \begin{subfigure}{0.45\linewidth}
        \centering
        \includegraphics[width=\linewidth]{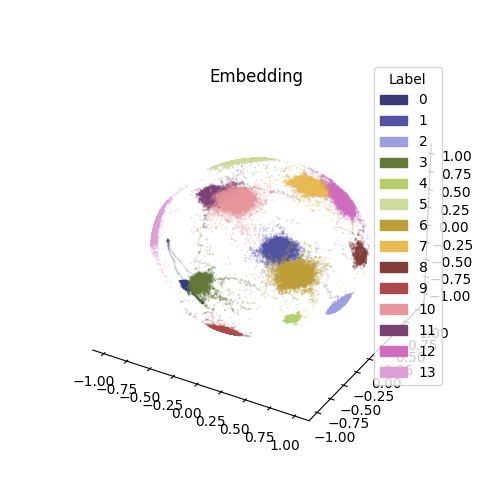}
        \caption{Normal embedding from V1 labels run}
    \end{subfigure}
    \hfill
    \begin{subfigure}{0.45\linewidth}
        \centering
        \includegraphics[width=\linewidth]{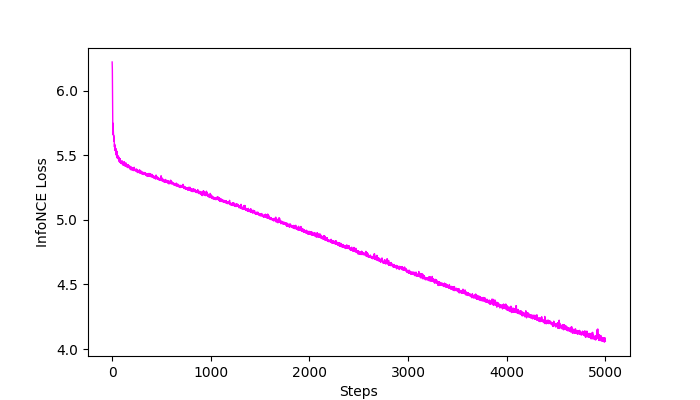}
        \caption{Normal embedding loss from V1 labels run}
    \end{subfigure}
        \hfill
    \begin{subfigure}{0.45\linewidth}
        \centering
        \includegraphics[width=\linewidth]{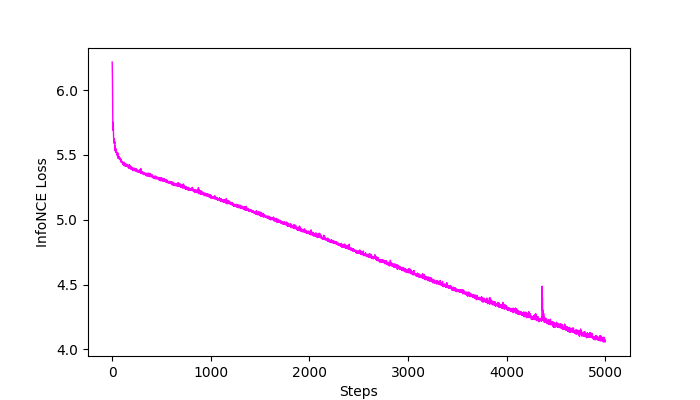}
        \caption{Pseudocut control loss from V1 labels run}
    \end{subfigure}
    \caption{Normal embedding and loss compared to pseudocut embedding and loss. Example V1 labels run0}
    \label{fig:v1_controls5}
\end{figure}

\textbf{V1: } This type of labeling assumed each interaction was uniquely labeled. So, within one dyad there were two different sets of labels, switching at the time speaker and listener changed their role. \\

\begin{tabular}{ccc}
\toprule
Dataset / Metrics & Goodness of Fit & KNN-5 Decoder Accuracy \\ \midrule
Normal data  (mean over 5 runs)      & 3.1342        & 0.99        \\ 
Behavioral control       & 0        & -1.374         \\ 
Neural control       & 0       & -1.383       \\ \
Pseudocut control       & 3.131        & 0.99        \\ \bottomrule
\end{tabular} \\

\begin{figure}[h]
    \centering
    \begin{subfigure}{0.45\linewidth}
        \centering
        \includegraphics[width=\linewidth]{combined_v1_run0_embedding.png}
        \caption{V1 Run 0 Embedding} 
    \end{subfigure}
    \hfill
    \begin{subfigure}{0.45\linewidth}
        \centering
        \includegraphics[width=\linewidth]{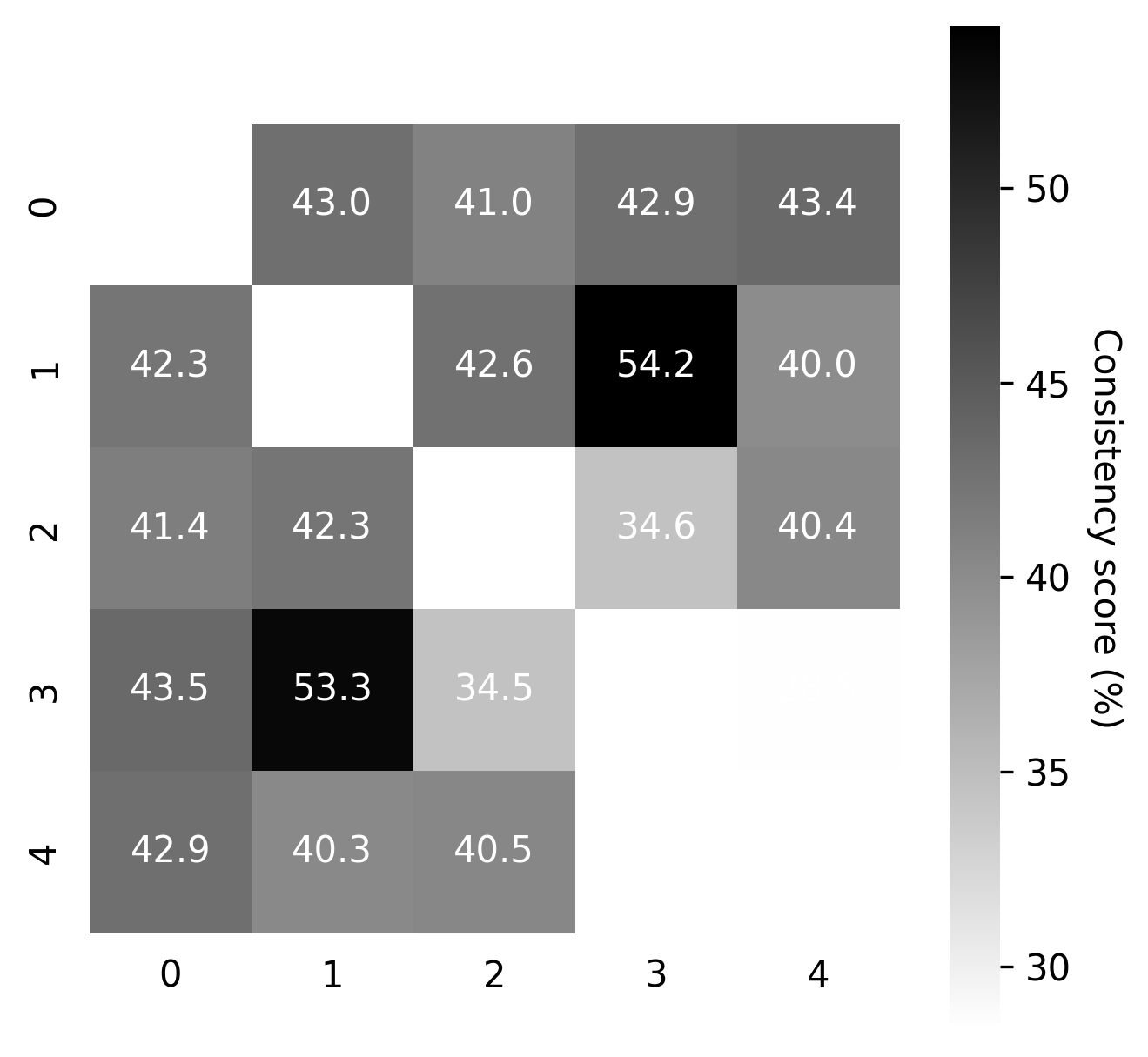}
        \caption{Consistency scores from V1}
    \end{subfigure}
    \caption{Results for V1}
    \label{fig:v1_controls2}
\end{figure}

\textbf{V2: } This type of labeling assumed each dyad received their own label. 

\begin{tabular}{ccc}
\toprule
Dataset / Metrics & Goodness of Fit & KNN-5 Decoder Accuracy \\ \midrule
Normal data  (mean over 5 runs)      & 2.6074        & 0.998       \\ 
Behavioral control       & 0        & -1.211         \\ 
Neural control       & 0       & -1.212       \\ 
Pseudocut control       & 2.594        & 0.99        \\ \bottomrule
\end{tabular} \\ 

\begin{figure}[h]
    \centering
    \begin{subfigure}{0.45\linewidth}
        \centering
        \includegraphics[width=\linewidth]{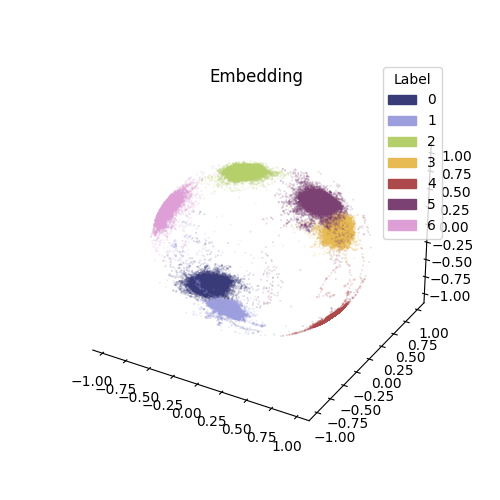}
        \caption{V2 Run 0 embedding} 
    \end{subfigure}
    \hfill
    \begin{subfigure}{0.45\linewidth}
        \centering
        \includegraphics[width=\linewidth]{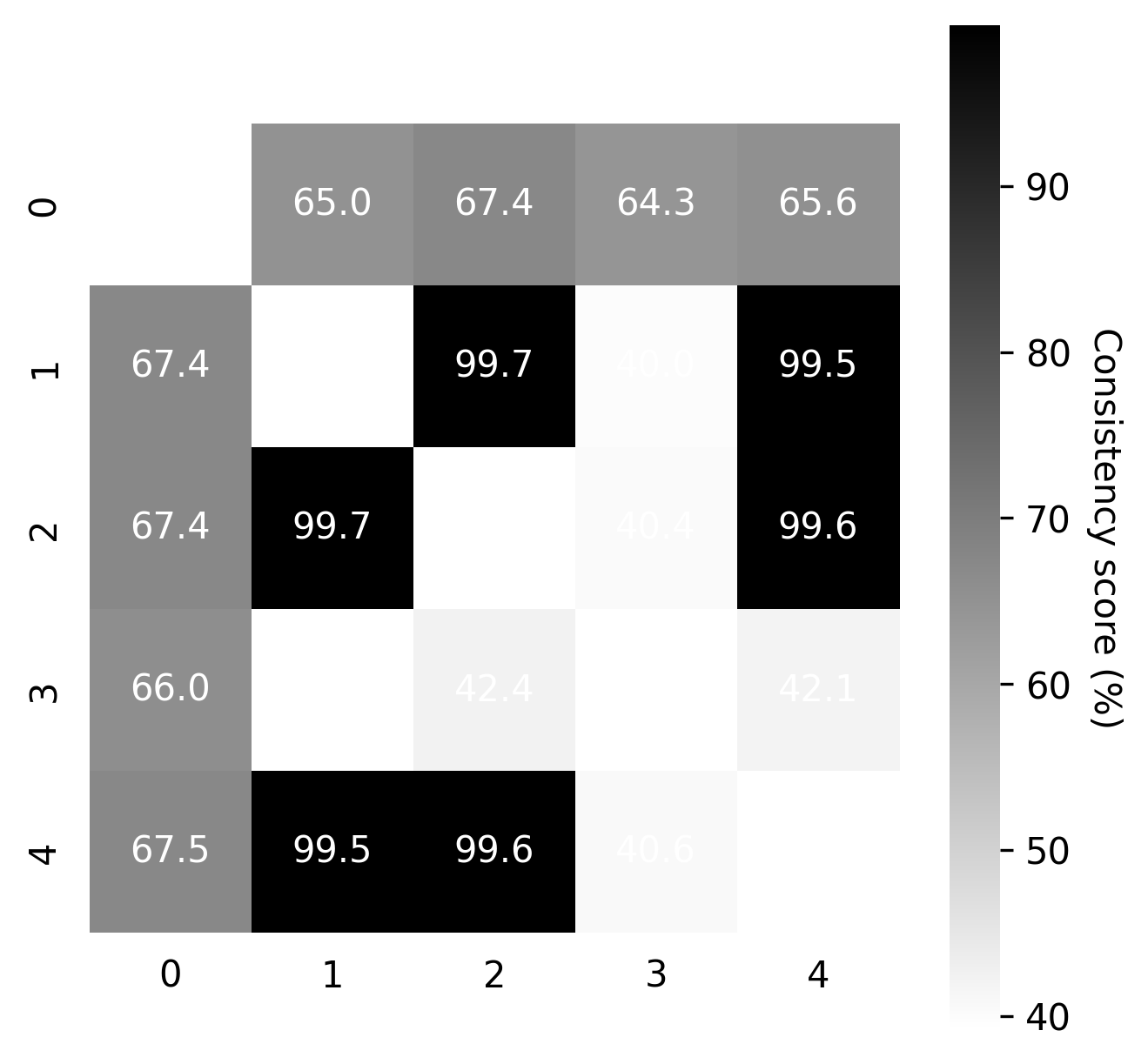}
        \caption{Consistency scores from V2}
    \end{subfigure}
    \caption{Results for V2}
    \label{fig:v1_controls3}
\end{figure} 

\textbf{V3: } This type of labeling was across pairs, the first part of the conversation was labelled as 0 and second part of the conversation was labeled as 1. 

\begin{tabular}{ccc}
\toprule
Dataset / Metrics & Goodness of Fit & KNN-5 Decoder Accuracy \\ \midrule
Normal data  (mean over 5 runs)      & 0.988        & 0.998       \\ 
Behavioral control       & 0        & -1.0         \\ 
Neural control       & 0       & -1.005       \\ 
Pseudocut control       & 0.988        & 0.998        \\ \bottomrule
\end{tabular} \\ 

\begin{figure}[h]
    \centering
    \begin{subfigure}{0.45\linewidth}
        \centering
        \includegraphics[width=\linewidth]{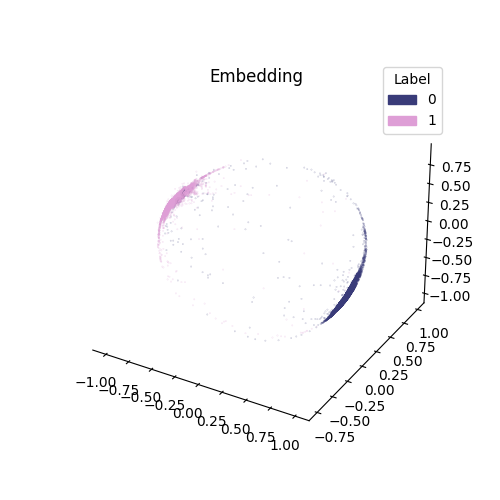}
        \caption{V3 Run 0 Embedding} 
    \end{subfigure}
    \hfill
    \begin{subfigure}{0.45\linewidth}
        \centering
        \includegraphics[width=\linewidth]{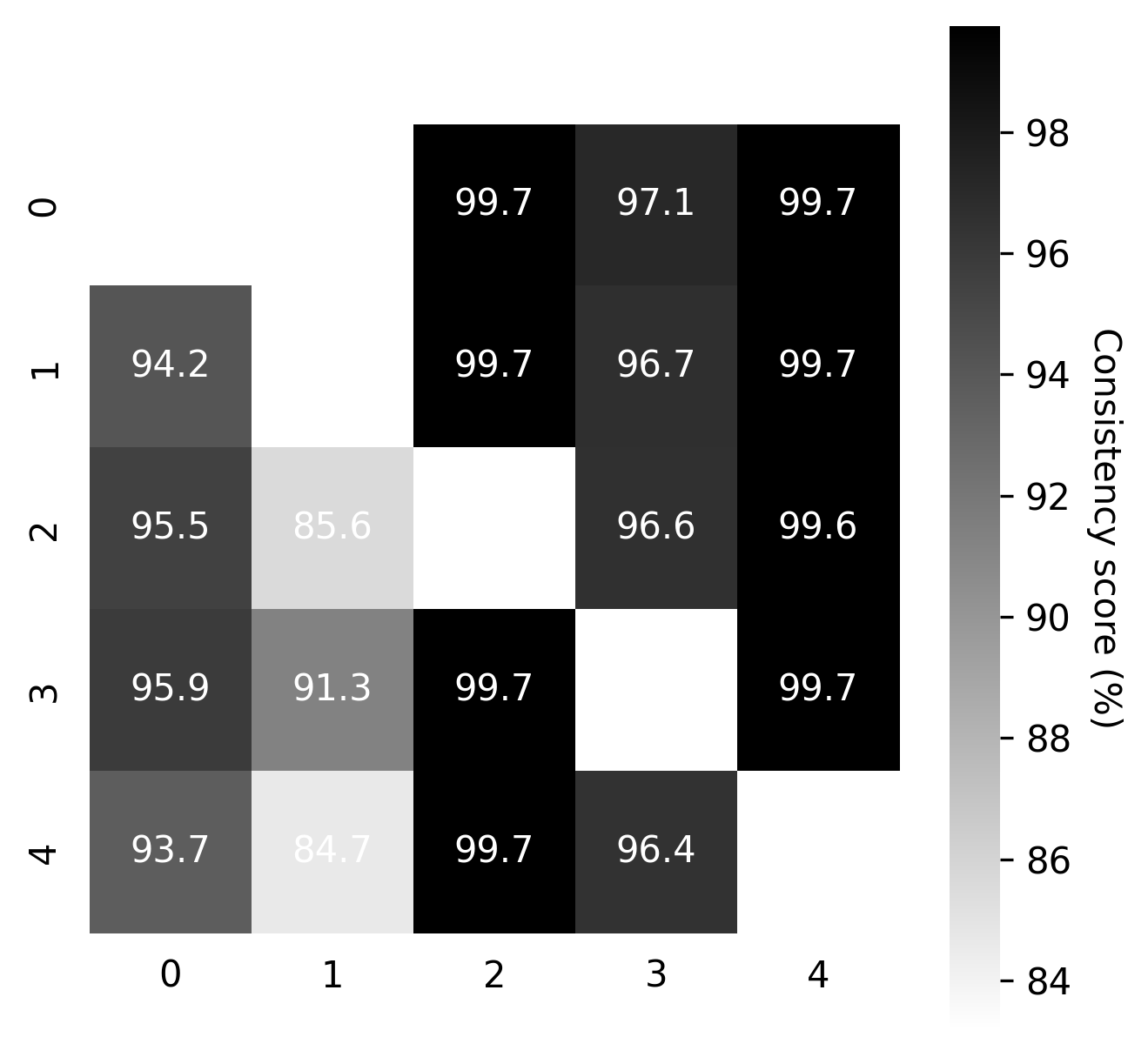}
        \caption{Consistency scores from V3}
    \end{subfigure}
    \caption{Results for V3}
    \label{fig:v1_controls4}
\end{figure} 

\textbf{V4: } This part of labeling gave each interaction the number corresponding to their differences in their AQ score as an absolute value. For the limited dataset, we have 4 different labels. 

\begin{tabular}{ccc}
\toprule
Dataset / Metrics & Goodness of Fit & KNN-5 Decoder Accuracy \\ \midrule 
Normal data  (mean over 5 runs)      & 1.8926        & 1.0       \\ 
Behavioral control       & 0        & -0.883         \\ 
Neural control       & 0       & -0.877       \\ 
Pseudocut control       & 1.891        & 1.0        \\ \bottomrule
\end{tabular} \\ 

\begin{figure}[h]
    \centering
    \begin{subfigure}{0.45\linewidth}
        \centering
        \includegraphics[width=\linewidth]{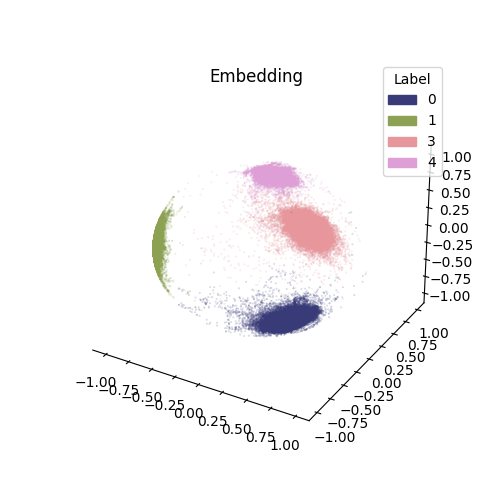}
        \caption{V4 Run 0 Embedding} 
    \end{subfigure}
    \hfill
    \begin{subfigure}{0.45\linewidth}
        \centering
        \includegraphics[width=\linewidth]{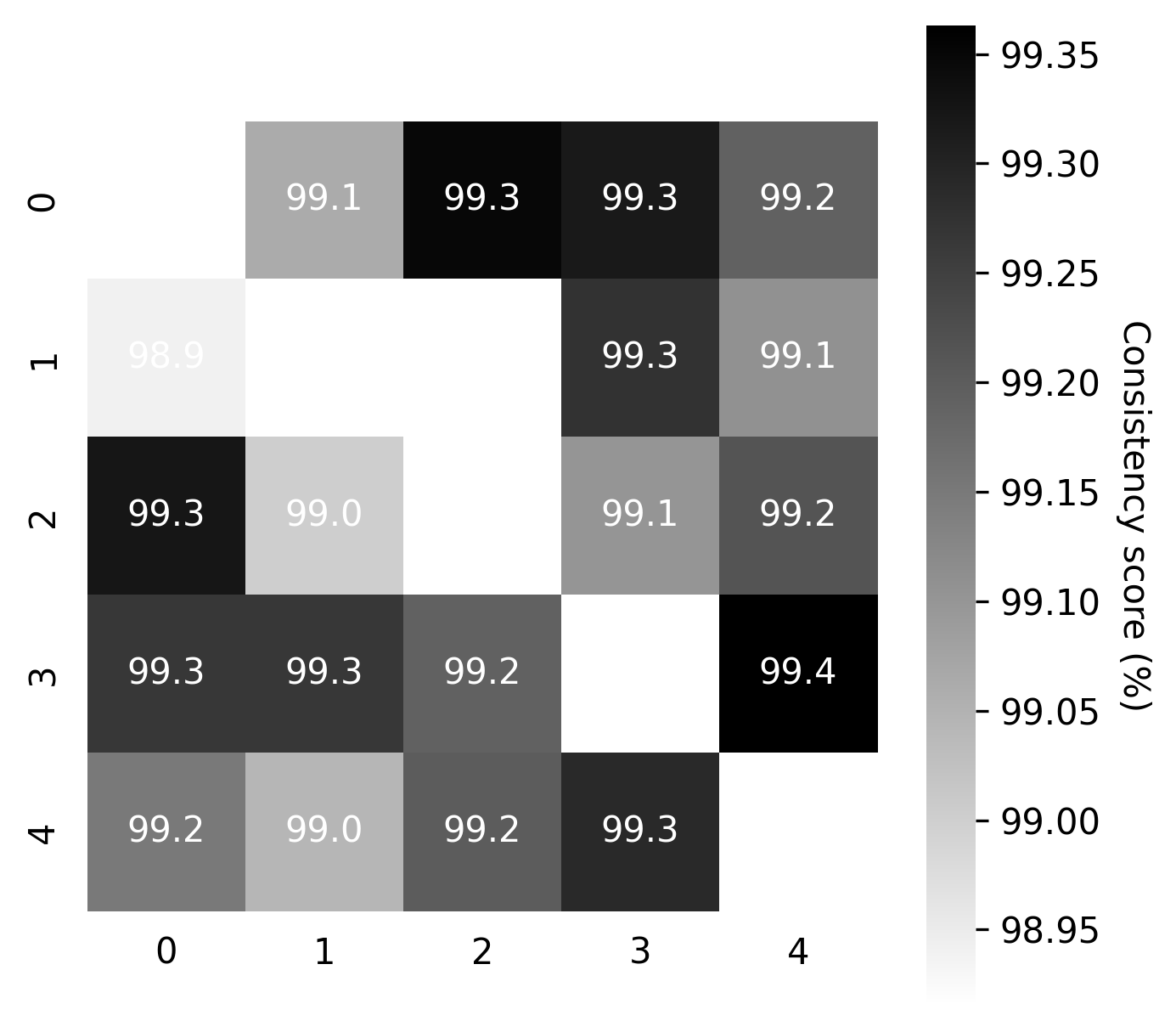}
        \caption{Consistency scores from V4}
    \end{subfigure}
    \caption{Results for V4}
    \label{fig:v1_controls}
\end{figure} 

\textbf{V5: } This type of labelling is identical to V4, except working with non-absolute labels. This means, that as participants were alternating in conversations, sometimes participants with a bigger or smaller AQ score spoke. However, if we subtract one value for another in different order, some values turn out to be negative, which CEBRA doesn't support in the current configuration. For that, we shifted the scale to positive numbers and had 7 unique labels. 

\begin{tabular}{ccc}
\toprule
Dataset / Metrics & Goodness of Fit & KNN-5 Decoder Accuracy \\ \midrule
Normal data  (mean over 5 runs)      & 2.442        & 1.0       \\ 
Behavioral control       & 0        & -0.883         \\ 
Neural control       & 0       & -0.877       \\ 
Pseudocut control       & 2.441        & 1.0        \\ \bottomrule
\end{tabular} \\ 

\begin{figure}[h]
    \centering
    \begin{subfigure}{0.45\linewidth}
        \centering
        \includegraphics[width=\linewidth]{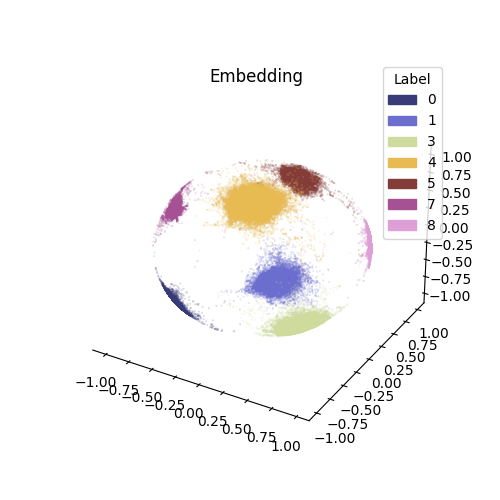}
        \caption{V5 Run 3 Embedding} 
    \end{subfigure}
    \hfill
    \begin{subfigure}{0.45\linewidth}
        \centering
        \includegraphics[width=\linewidth]{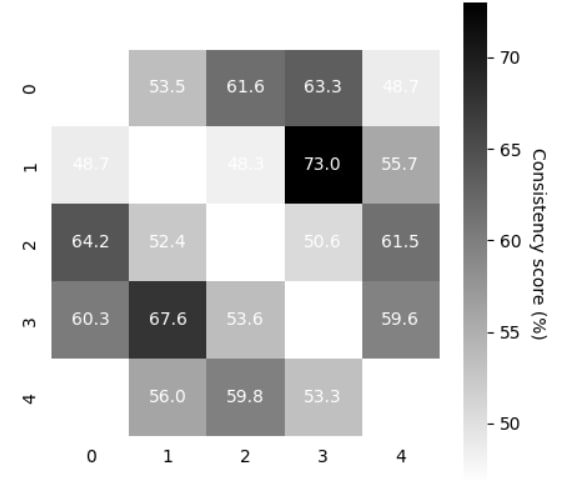}
        \caption{Consistency scores from V5}
    \end{subfigure}
    \caption{Results for V5}
    \label{fig:v1_controls6}
\end{figure}

\textbf{V6: } This version has been designed as a sanity check. As the best goodness-of-fit was achieved through the smallest partition of the data, we thought that maybe the algorithm might just produce the best goodness-of-fit in case of smaller partitioning. To test that, we ran a version with 30 equally spaced random labels. The mean goodness of fit is: 2.71

\textbf{N+1 (leave-one-dyad-out) hold-out and gender coding. } 
We evaluated cross-dyad generalization by training CEBRA on all dyads except one and testing only on the held-out dyad (label version V3: direction label \emph{0} = first role block, \emph{1} = second role block after role swap). We also coded dyad gender per direction (codes: 1=FF, 2=MM, 3=SpF–ListM, 4=SpM–ListF) and summarize performance by gender grouping. Across five seeds per hold-out, \textbf{same-gender (FF) dyads} showed higher and more stable decoding than \textbf{mixed-gender dyads}. Specifically, FF dyads achieved mean KNN accuracy $0.956 \pm 0.049$ with GOF $0.314 \pm 0.319$, whereas mixed dyads achieved $0.854 \pm 0.179$ with GOF $0.126 \pm 0.491$. Notably, two mixed dyads (nt38–39, nt36–37) were consistently harder (KNN $\approx 0.70$–$0.73$), while nt40–41 generalized strongly (KNN $\approx 0.996$).
\\[2mm]

\begin{center}
\begin{tabular}{ccccc}
\toprule
Dyad group & Runs & \#Dyads & GOF (mean $\pm$ sd) & KNN (mean $\pm$ sd) \\ \midrule
FF & 15 & 3 & 0.314 $\pm$ 0.319 & 0.956 $\pm$ 0.049 \\ 
Mixed & 20 & 4 & 0.126 $\pm$ 0.491 & 0.854 $\pm$ 0.179 \\ \bottomrule
\end{tabular}
\end{center}
\vspace{2mm}

\noindent\small\textit{Per-dyad N+1 summary (means over 5 runs). Gender codes (dir0–dir1) in filenames: 1=FF, 2=MM, 3=SpF–ListM, 4=SpM–ListF.}
\begin{center}
\small
\begin{tabular}{ccccc}
\toprule
Dyad (held-out) & Gender group & GOF (mean $\pm$ sd) & KNN (mean $\pm$ sd) & Eval samples \\ \midrule
nt9--10 (g3--4)  & Mixed & $-0.018 \pm 0.385$ & $0.995 \pm 0.004$ & 105{,}250 \\ 
nt34--35 (g1--1) & FF    & $0.323 \pm 0.272$  & $0.988 \pm 0.010$ & 74{,}500  \\ 
nt21--22 (g1--1) & FF    & $0.384 \pm 0.411$  & $0.948 \pm 0.053$ & 80{,}750  \\ 
nt38--39 (g4--3) & Mixed & $-0.043 \pm 0.403$ & $0.700 \pm 0.159$ & 64{,}750  \\ 
nt40--41 (g4--3) & Mixed & $0.740 \pm 0.415$  & $0.996 \pm 0.002$ & 79{,}750  \\ 
nt36--37 (g4--3) & Mixed & $-0.173 \pm 0.128$ & $0.725 \pm 0.162$ & 84{,}250  \\ 
nt13--14 (g1--1) & FF    & $0.235 \pm 0.316$  & $0.931 \pm 0.059$ & 129{,}750 \\ \bottomrule
\end{tabular}
\end{center}
\normalsize
\vspace{2mm}

\begin{figure}[h]
    \centering
    \begin{subfigure}{0.45\linewidth}
        \centering
        \includegraphics[width=\linewidth]{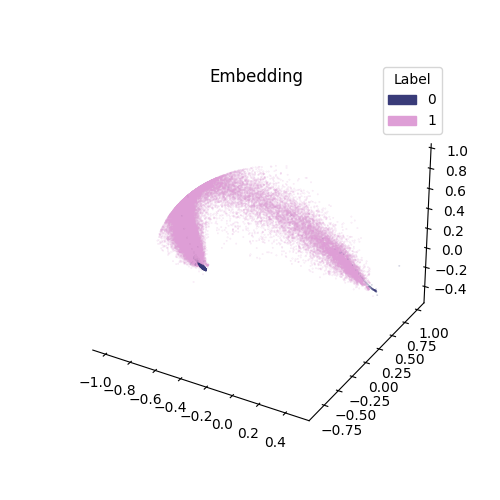}
        \caption{Hold-out: nt9--10 (Mixed, g3--4)}
    \end{subfigure}
    \hfill
    \begin{subfigure}{0.45\linewidth}
        \centering
        \includegraphics[width=\linewidth]{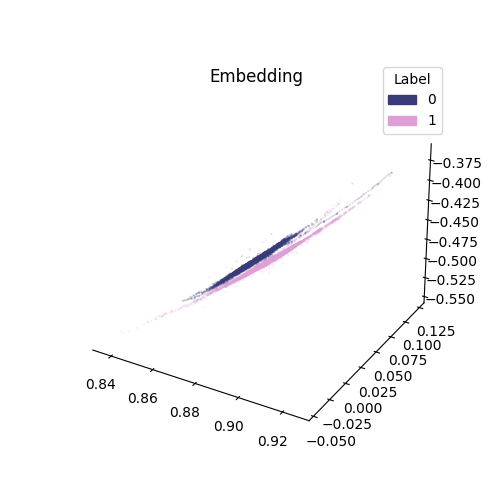}
        \caption{Hold-out: nt34--35 (FF, g1--1)}
    \end{subfigure}
    \hfill
    \begin{subfigure}{0.45\linewidth}
        \centering
        \includegraphics[width=\linewidth]{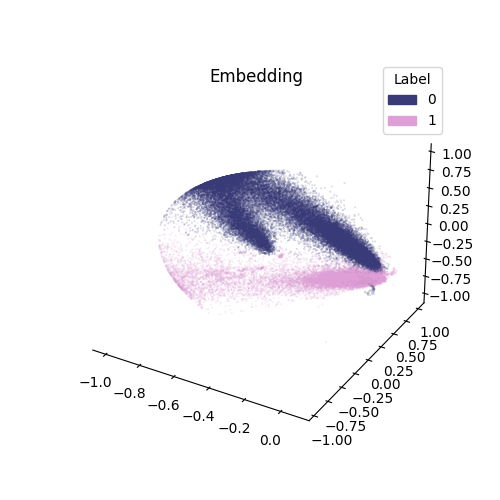}
        \caption{Hold-out: nt21--22 (FF, g1--1)}
    \end{subfigure}
    \hfill
    \begin{subfigure}{0.45\linewidth}
        \centering
        \includegraphics[width=\linewidth]{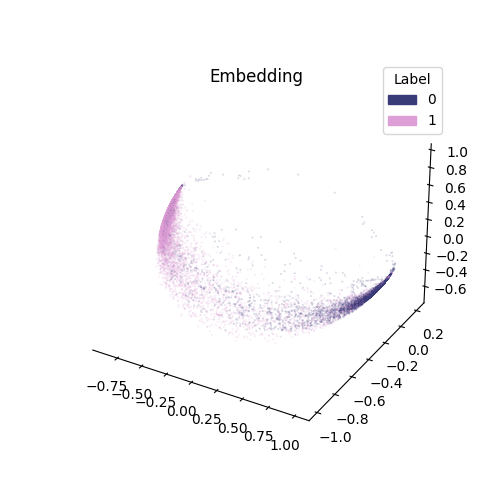}
        \caption{Hold-out: nt38--39 (Mixed, g4--3)}
    \end{subfigure}
    \caption{\textbf{N+1 embeddings (run 4), part 1.} Points colored by V3 direction labels (\textbf{0} = first role block; \textbf{1} = second block after role swap). Legends in each panel match training labels; filenames include held-out dyad and gender codes (g\{dir0\}--g\{dir1\}).}
    \label{fig:nplus1_embeds_a}
\end{figure}

\begin{figure}[h]
    \centering
    \begin{subfigure}{0.45\linewidth}
        \centering
        \includegraphics[width=\linewidth]{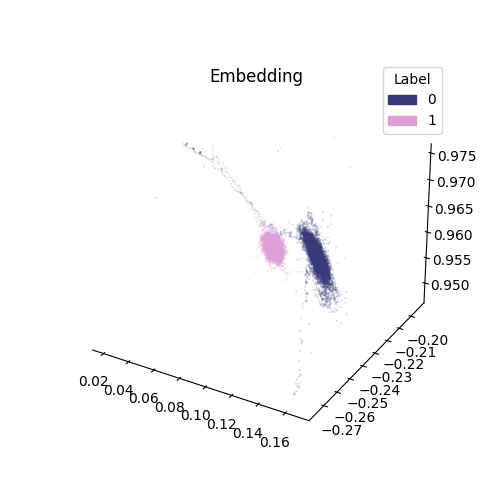}
        \caption{Hold-out: nt40--41 (Mixed, g4--3)}
    \end{subfigure}
    \hfill
    \begin{subfigure}{0.45\linewidth}
        \centering
        \includegraphics[width=\linewidth]{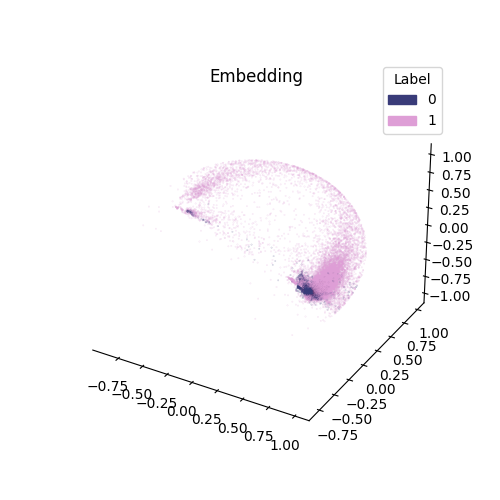}
        \caption{Hold-out: nt36--37 (Mixed, g4--3)}
    \end{subfigure}
    \hfill
    \begin{subfigure}{0.45\linewidth}
        \centering
        \includegraphics[width=\linewidth]{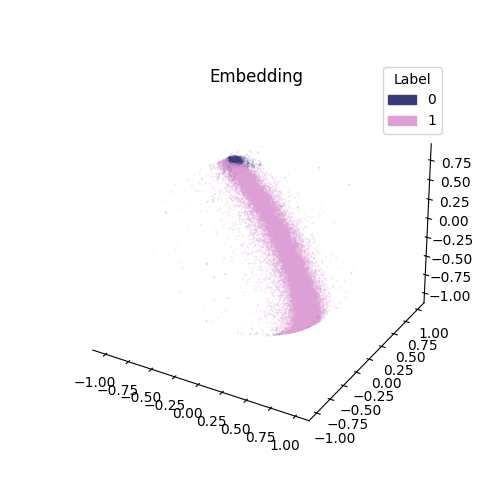}
        \caption{Hold-out: nt13--14 (FF, g1--1)}
    \end{subfigure}
    \caption{\textbf{N+1 embeddings (run 4), part 2.} Same conventions as Fig.~\ref{fig:nplus1_embeds_a}.}
    \label{fig:nplus1_embeds_b}
\end{figure}

\textbf{Ablation analysis} In order to check what these representations might refer to in the domain of congitive processes we performed ablation analysis. Meaning, we trained the models with several channels being ablated depending on the version. In V3 channels were ablated that were thought to be responsible for listening comprehension and speech production, while in V5 channels that were associated with executive functioning were ablated. 

The mean goodness-of-fit for V3 ablation: 0.979 

The mean goodness-of-fit for V5 ablation is: 2.442 

Both of them not diverging significantly from the non-ablated results (refer to V3 and V5 respective analysis)

\section{Discussion}

We found embeddings with well-separated classes using CEBRA for all behavioral labels, e.g., listener-speaker roles, differences in AQ score, etc. To ensure interpretability of these results, we developed several types of control analyses. The primary validation (randomly shuffling neural data and observation labels separately) reduced goodness-of-fit and classification accuracy, confirming that our embeddings capture relationships between neural activity and behavioral variables. We observed no decrease in goodness-of-fit and classification accuracy after introducing random cuts to the data (making it discontinuous). This suggests that cleaning processes involving segment removal do not influence the embedding and classification.
The ablation analysis, where we removed channels associated with specific cognitive processes (listening comprehension/speech production for V3, ASD-related channels for V5), produced results that did not clearly differentiate these functional domains as expected. This likely indicates that neural representations in social interaction are distributed broadly, and not limited to speaking and listening-specific neural regions.
Despite having better goodness-of-fit with finer-grained labeling schemes, we observed lower consistency between embeddings across different runs. This trade-off between model fit and stability suggests that while CEBRA can capture detailed individual differences, the most generalizable findings are based on broad categorical distinctions (such as speaker vs. listener roles). The cross-dyad generalization analysis (N+1 leave-one-dyad-out) supported this finding: same-gender dyads showed more stable decoding than mixed-gender dyads, and certain dyads were consistently more difficult to generalize to, suggesting that some aspects of dyadic neural coordination may be modulated by participant characteristics beyond the ones we labeled.
While our approach demonstrates that CEBRA can extract meaningful representations from hyperscanning data, the emphasis on individual variability suggests that generalizable models may require substantially larger and more diverse datasets, or may need to incorporate explicit modeling of individual differences.

\textbf{Limitations}: Our sample size is relatively small (N=8 dyads), within a narrow range of AQ scores, which limits conclusion generalizability: we demonstrate proof-of-concept for CEBRA's application to hyperscanning data, but larger samples would be needed to establish clinical or diagnostic applications. 

\textbf{Broader Impacts}: Our approach advances understanding of interpersonal neural mechanisms in autism and social communication more broadly. By moving beyond deficit-based models to examine dyadic interactions, this research supports more nuanced views of neurodiversity and could inform more effective, relationship-focused interventions. The method could contribute to development of objective biomarkers for social communication challenges. Our work aims to understand natural variations in social communication rather than pathologize them, but it is important to prevent misapplication of the method. By better understanding how different neurotypes communicate most effectively, we may be able to design more inclusive social environments and technologies.

 \section*{References}






\section*{Acknowledgments}

 MG and GH were participants in the 2025 Google Summer of Code Program with Machine Learning for Science (ML4SCI) Organization.

Data collection for this research was supported by funding to EM from the University of Alabama College of Arts and Sciences.

\end{document}